\documentclass[
reprint,
superscriptaddress,
amsmath,amssymb,
aps,
prx,
]{revtex4-2}

\usepackage{import}
\usepackage{graphicx}% Include figure files
\usepackage{dcolumn}% Align table columns on decimal point
\usepackage{bm}% bold math
\usepackage[utf8]{inputenc}
\usepackage[T1]{fontenc}
\usepackage{mathptmx}
\usepackage{etoolbox}
\usepackage[colorlinks=true, allcolors=blue]{hyperref}

\newcommand{\TiOtwo}{TiO\textsubscript{2}}

\begin{document}

\preprint{APS/123-QED}

% \title[]{Long-lived Optical Coherence in Erbium-based Quantum Memory Platform via Back End of Line Deposition on Foundry Photonic Integrated Circuits} 

% \title[]{High Optical Coherence in Erbium-doped Thin Films Integrated with Foundry Photonic Circuits  via Backend Deposition for Quantum Memories} 

% \title[]{High Optical Coherence in Erbium Quantum Memory Platform Integrated via Backend Deposition with Foundry Photonic Circuits} 

\title[]{Erbium Quantum Memory Platform with Long Optical Coherence via Back-End of Line Deposition on Foundry-Fabricated Photonics}

% Tentative author list\\
\author{Shobhit Gupta}
\email{shobhit@memq.tech}
 \affiliation{memQ, Inc, Chicago, IL 60615, USA}
\author{Robert M. Pettit}
 \affiliation{memQ, Inc, Chicago, IL 60615, USA}

\author{Ananthesh Sundaresh}
 \affiliation{memQ, Inc, Chicago, IL 60615, USA}

  \author{Vasileios Niaouris}%
\affiliation{Argonne National Laboratory, Lemont, Illinois 60439, USA}%

 \author{ Skylar Deckoff-Jones}
 \affiliation{memQ, Inc, Chicago, IL 60615, USA}
 
  \author{Daniel P. Crowley}
 \affiliation{American Institute for Manufacturing Integrated Photonics (AIM Photonics), Albany, NY 12203, USA,}

  \author{Lewis G. Carpenter}
 \affiliation{American Institute for Manufacturing Integrated Photonics (AIM Photonics), Albany, NY 12203, USA,}

\author{Alan M. Dibos}%
\affiliation{Argonne National Laboratory, Lemont, Illinois 60439, USA}%

 \author{Manish Kumar Singh}
 \affiliation{memQ, Inc, Chicago, IL 60615, USA} 
 
\author{Sean E. Sullivan}
 \email{sean@memq.tech}
 \affiliation{memQ, Inc, Chicago, IL 60615, USA}

\date{\today}% It is always \today, today,

\begin{abstract}
Realizing scalable quantum interconnects necessitates the integration of solid-state quantum memories with foundry photonics processes. While prior photonic integration efforts have relied upon specialized, laboratory-scale fabrication techniques, this work demonstrates the monolithic integration of a quantum memory platform with low-loss foundry photonic circuits via back-end-of-line deposition. We deposited thin films of titanium dioxide (\TiOtwo) doped with erbium (Er) onto silicon nitride nanophotonic waveguides and studied Er optical coherence at sub-Kelvin temperatures with photon echo techniques. We suppressed optical dephasing through ex-situ oxygen annealing and optimized measurement conditions, which yielded an optical coherence time of 64~$\mu$s (a 5~kHz homogeneous linewidth) and slow spectral diffusion of 27~kHz over 4~ms, results that are comparable to state-of-the-art erbium devices. Combined with second-long electron spin lifetimes and demonstrated electrical control of Er emission, our findings establish Er:\TiOtwo\ on foundry photonics as a manufacturable platform for ensemble and single-ion quantum memories. 

\end{abstract}

\maketitle

\section{Introduction}
Rare-earth ions (REI) doped in solids are leading candidates for quantum memories due to their exceptional optical and spin coherence times in a solid-state medium \cite{THIEL2011353,Rancic2018,Zhong2015}. The trivalent erbium ion (Er$^{3+}$ but referred to as Er for brevity) in particular, is interesting for quantum networking \cite{Kimble2008} due to its emission in the low-loss telecom C-band and its long spin coherence times at sub-Kelvin temperatures \cite{Raha2020, Asadi2020,PhysRevApplied.19.044029,doi:10.1126/sciadv.abj9786}. Scalable quantum networking would require chip-scale integration of quantum memories with photonic control elements, including modulators, filters, switches, and detectors \cite{https://doi.org/10.1002/lpor.202300257,PRXQuantum.5.010101,10.1063/5.0179539}. Although photonic integrated circuits can be mass-produced using commercial foundry processes, the integration of rare-earth ion quantum memories with photonic foundry processes remains an outstanding challenge \cite{Becher_2023,10019546,10.1063/5.0156874}. 
State-of-the-art REI quantum memories are primarily based on well-studied bulk crystal hosts such as Y\textsubscript{2}SiO\textsubscript{5} \cite{afzelius2,Saglamyurek2011,PhysRevApplied.5.054013,PhysRevLett.125.260504,PhysRevLett.129.210501}, YVO\textsubscript{4} \cite{PhysRevB.98.024404,Liang:20}, and LiNbO\textsubscript{3} \cite{doi:10.1126/sciadv.adf4587,Zhang:24}. Limited photonic integration has been demonstrated in bulk crystals using specialized laboratory fabrication techniques which involve either structuring the rare-earth host medium \cite{PhysRevApplied.12.024062,Saglamyurek2011,PhysRevApplied.5.054013,Zhang:24, PhysRevLett.125.260504,PhysRevLett.129.210501,doi:10.1126/sciadv.adf4587,doi:10.1021/acs.nanolett.9b04679,PhysRevApplied.18.014069}  or creating heterogeneous photonic structures on top of the host crystal \cite{Okajima:2025lxc,Yang:21,Miyazono:17,Craiciu:21,doi:10.1126/sciadv.adu0976}. 

% Photonic integration efforts for bulk crystals have been based on structuring the crystal medium with focused ion beam milling (FIB)\cite{PhysRevApplied.12.024062}, titanium indiffusion \cite{Saglamyurek2011}, femtosecond laser machining \cite{PhysRevApplied.5.054013,Zhang:24, PhysRevLett.125.260504,PhysRevLett.129.210501,doi:10.1126/sciadv.adf4587}and SmartCut \cite{doi:10.1021/acs.nanolett.9b04679,PhysRevApplied.18.014069} techniques or creating heterogeneous photonic structures on top of the REI host crystal by transfer printing\cite{Okajima:2025lxc}, flip-chip bonding\cite{Yang:21}, or photonic layer deposition\cite{Miyazono:17,Craiciu:21}. Although these works have reduced the quantum memory footprint, their scalability and photonic integration prospects are ultimately limited by the bulk crystals.  

\begin{figure*}[!t]
    \centering
    \includegraphics[width=0.99\linewidth]{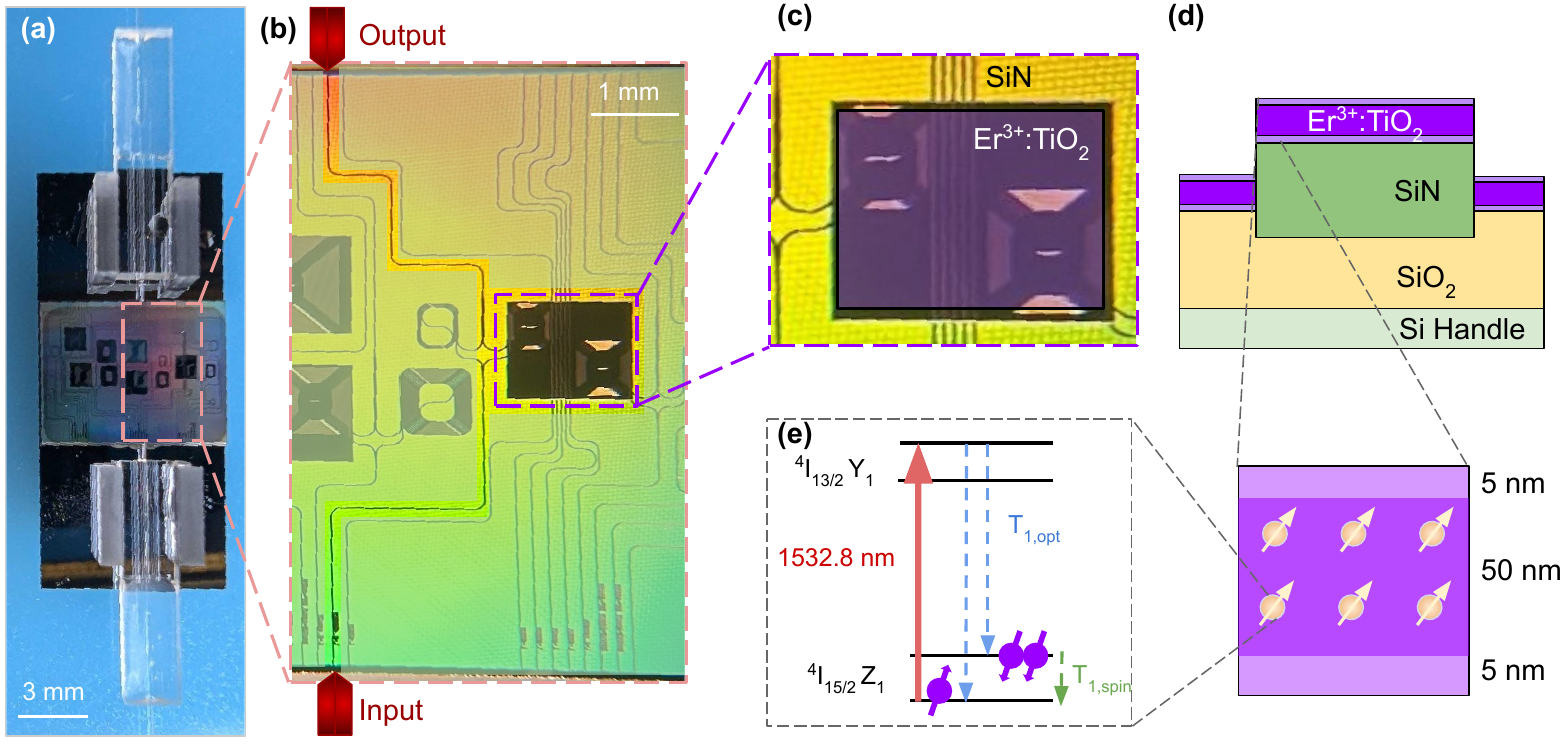}
    \caption{ Er-doped thin-film integrated on a fully-packaged foundry photonic chip. (a) Packaged device with input/output lensed fibers permanently mounted on a carrier chip. (b) Image of the photonic chip where the particular waveguide used in the measurement has been highlighted for clarity. (c) Zoomed-in image of an arm of the MZI showing the unclad window used for \TiOtwo\ deposition (purple). (d, top) Cross-section schematic of the device material stack. (d, bottom) A zoom-in of the 60~nm thick \TiOtwo\ film deposited on the unclad silicon nitride waveguide, with a 50~nm thick Er-doped layer sandwiched between 5~nm undoped layers.  (e) Energy level structure of the $\mathrm{^{4}I_{15/2}Z_1 \, \rightarrow ^{4}I_{13/2}Y_1}$ transition for Er$\mathrm{^{3+}}$ in an applied magnetic field.}
    \label{fig1}
\end{figure*}

The challenges of scalable photonic integration in bulk crystals have necessitated the development of high-quality, quantum-grade rare-earth ion doped thin films grown at the wafer scale on a silicon-on-insulator (SOI) platform \cite{10.1063/1.5142611,doi:10.1021/acs.nanolett.2c01561,10.1063/5.0176610,10.1063/5.0224010,10.1063/5.0010833,doi:10.1021/acs.jpcc.9b02597,10.1063/5.0222269}. 
Unlike conventional photonic integration where thin film growth precedes device fabrication \cite{10.1063/5.0222269,doi:10.1021/acs.nanolett.2c01561,doi:10.1021/acs.nanolett.2c01561}, this work investigates back-end-of-line (BEOL) integration, where thin films are grown directly onto prefabricated photonic structures.  This facilitates seamless integration of the REI quantum memory layer with commercial foundry processes, enabling independent manufacturing of photonic elements and subsequent memory qubit layer integration. Using BEOL integration, our previous work demonstrated Purcell enhancement of single Er emitters in polycrystalline \TiOtwo\ thin films using nanophotonic cavities on a 300-mm silicon foundry platform \cite{pettit2025monolithicallyintegratedcbandquantum}. 

A significant challenge in integrating emitters with nanophotonic structures arises from their required proximity to material interfaces, which show increased structural disorder compared to the bulk (attributed to tunneling two-level systems (TLS) \cite{MACFARLANE1987179}), ultimately resulting in degraded emitter coherence \cite{Becher_2023,PhysRevApplied.20.044018,PhysRevX.10.041025}. This challenge is exacerbated for emitters within polycrystalline thin films, where the emitters are in proximity to both grain boundaries and heterogeneous material interfaces. Consequently, their application in quantum memories depends on the critical tasks of identifying and quantifying dephasing mechanisms, followed by the development of robust noise suppression strategies, including but not limited to annealing \cite{PhysRevB.105.224106}, optimizing measurement conditions  \cite{PhysRevApplied.19.044029,prbsubkhz}, and symmetry-based reduced sensitivity to noise \cite{Ourari2023,horvath2023strongpurcellenhancementoptical}. 
 
We investigated Er-doped titanium dioxide (\TiOtwo) polycrystalline thin films grown using molecular beam deposition on foundry-fabricated silicon nitride nanophotonic waveguides. \TiOtwo\ offers compatibility with complementary metal-oxide-semiconductor (CMOS) processing, a low magnetic noise environment and non-polar symmetry, making it uniquely suitable for photonic integration of rare-earth ions \cite{10019546,doi:10.1021/acs.nanolett.9b03831,10019546,pettit2025monolithicallyintegratedcbandquantum}. Single Er emitters in \TiOtwo\ thin films have recently been integrated with nanophotonic cavities \cite{10.1063/5.0222269,pettit2025monolithicallyintegratedcbandquantum}; however, the optical and spin coherence properties of Er in the \TiOtwo\ host are largely unknown. Using photon echo techniques, we perform a high-resolution probe of optical homogeneous linewidth broadening mechanisms in waveguide-coupled Er ions at sub-Kelvin temperatures. Analysis of optical dephasing and spectral diffusion revealed characteristics indicative of a paramagnetic spin bath associated with oxygen-deficiency defects as well as TLS defects. Through optimization of the measurement conditions and extended oxygen annealing, we show suppression of both dephasing and spectral diffusion by an order of magnitude, leading to optical coherence times up to 64~\textmu s and a spectral diffusion linewidth down to 27~kHz over 4~ms. We also show that this platform offers electrical control over Er emission and long electron spin lifetimes, both essential for quantum memory applications. Our findings also highlight the viability of scalable foundry photonic integration while exhibiting high emitter coherence.

\begin{figure*}[!t]
    \centering
    \includegraphics[width=0.99\linewidth]{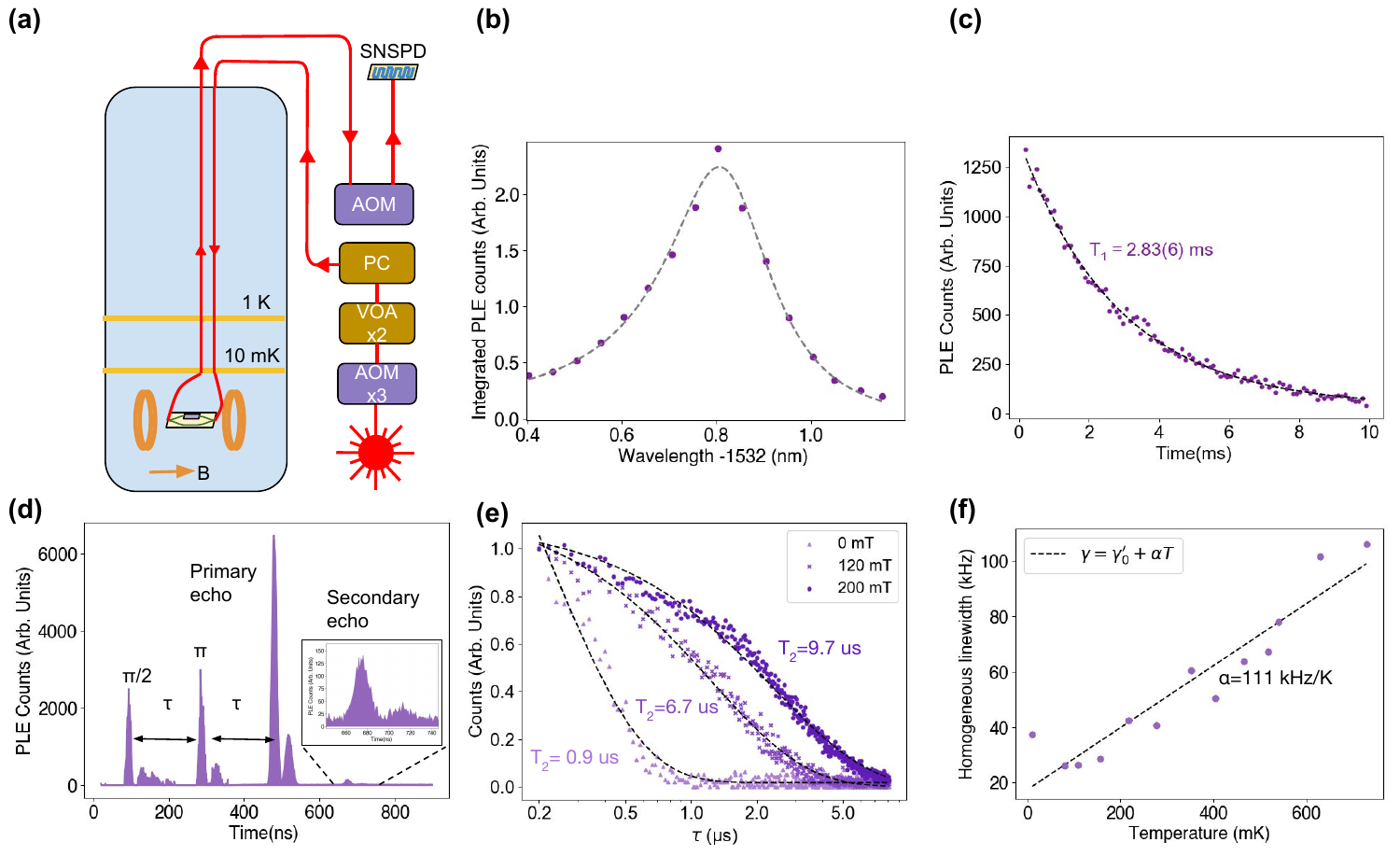}
    \caption{ Coherent optical spectroscopy at sub-Kelvin temperatures. (a) Experimental setup with only one magnetic axis shown for clarity, VOA: Variable optical attenuator, PC: Polarization controller (b) PLE scan around the expected Y\textsubscript{1}-Z\textsubscript{1} peak of Er in anatase phase fitted with a Lorentzian with FWHM linewidth of 0.28~nm (36~GHz) (c)  PLE decay corresponding to the 1532.8~nm peak (d) Photon echo time trace showing $\pi/2$, $\pi$ pulses (attenuated by $\sim$ 40~dB) separated by time $\tau$, followed by primary at time $\mathrm{2\tau}$ and secondary echo  at $3\tau$ (e) Photon echo decay corresponding to $B=0$~mT, 120~mT, and 200~mT (f) Photon echo homogeneous linewidth at 200~mT vs stage temperature fit with a linear broadening term attributed to TLS.}
    \label{fig2}
\end{figure*}
\section{Experimental Methods}

Our device platform is shown in Fig.~\ref{fig1}, and consists of a silicon nitride-based waveguide chip produced at AIM Photonics with back-end-of-the-line Er-doped \TiOtwo\ deposition and has been fully packaged with lensed optical fibers, as can be seen in Fig.~\ref{fig1}(a). This particular chip was fabricated in a multi-project wafer (MPW) run, and it has several single-mode silicon nitride nanophotonic waveguides with varying lengths and a thick SiO\textsubscript{2} cladding layer. The waveguide used in this measurement has been highlighted in Fig.~\ref{fig1}(b), and comprises a Mach-Zehnder interferometer (MZI) with the arms of the MZI containing a long spiral waveguide with a length of 0.486~cm. One arm of the MZI had the SiO\textsubscript{2} cladding layer removed during the MPW processing to leave a window (purple dashed line) where \TiOtwo\ could be deposited directly on the waveguide using molecular beam deposition, as shown in Fig.~\ref{fig1}(c) (Supplemental Material \cite{suppl}). We also note that since the film was only deposited on one arm of the MZI, we do not anticipate the MZI interference to affect the measurements and the MZI provides a fixed 3~dB loss (see Supplemental Material for device transmission \cite{suppl}). The choice of the MZI device for film deposition was motivated by its availability in the MPW run and its utility in estimating the refractive index of the \TiOtwo\ film. We performed the doped \TiOtwo\ growth onto two identically fabricated chips, and we performed ex-situ oxygen annealing for one hour and three hours, to produce chip1h and chip3h, respectively. All measurements in the manuscript, except those shown in Fig.~3 of the main text, were performed exclusively on the chip1h sample. 

Fig.~\ref{fig1}(d, top) shows the cross-sectional view of the material stack consisting of an Er-doped \TiOtwo\ layer (purple) deposited on top of silicon nitride nanophotonic waveguides (green). The 60~nm thick \TiOtwo\ layer comprises an undoped capping layer of 5~nm at the top and bottom and a 50~nm Er-doped layer, Fig.~\ref{fig1}(d, bottom) with a nominal density of 50 parts per million (ppm) in the middle \cite{10.1063/5.0224010}. The film is mixed-phase and polycrystalline \cite{10.1063/5.0176610} with the anatase phase of \TiOtwo\ (grain size $\sim$ 13.5(3) nm) being the focus of this work. The transverse electric (TE) mode of the nanophotonic waveguide is evanescently coupled to the Er-doped \TiOtwo\ layer at the top (Supplemental Material \cite{suppl}). Fig.~\ref{fig1}(e) shows the 1532~nm transition between the crystal field sublevels $\mathrm{^{4}I_{15/2}Z_1 \, \rightarrow ^{4}I_{13/2}Y_1}$  for Er$\mathrm{^{3+}}$ in \TiOtwo\ in red arrows. The optical transition has a lifetime of $\sim$ milliseconds, while the ground state level $\mathrm{^{4}I_{15/2}Z_1}$ has effective spin $\mathrm{1/2}$ at cryogenic temperature, with spin lifetimes generally between milliseconds to seconds \cite{PhysRevApplied.19.044029}.

\section{Optical spectroscopy}

\subsection{Photoluminescence excitation (PLE)}
The experimental setup is shown in Fig.~\ref{fig2}(a) with the sample mounted on a cold finger attached to the mixing chamber stage of an optical-fiber-equipped Bluefors dilution refrigerator with a nominal base temperature of 10~mK. A magnetic field was applied using a three-axis vector magnet up to 250~mT on each axis. Laser light was pulsed using three acousto-optical modulators (AOMs) in series, followed by variable optical attenuators and a fiber polarization controller. Another AOM was used in the collection path after the chip to shield the SNSPD during the excitation pulses to prevent latching. Additional details about the setup are provided in the Supplemental Material \cite{suppl}. 

Photoluminescence excitation (PLE) measurements were performed by sending optical pulses of 100~ns length and 10~nW power into the cryostat. Fig.~\ref{fig2}(b) displays the peak attributed to the Y\textsubscript{1} - Z\textsubscript{1} transition of Er\textsuperscript{3+} in anatase \TiOtwo . The peak was fitted with a Lorentzian with a FWHM linewidth of 0.28~nm. Fig.~\ref{fig2}(c) shows the PLE decay at a wavelength of 1532.8~nm with a lifetime ($T_{1,opt}$) of 2.8~ms.

\subsection{Photon echo homogeneous linewidth}

Photon echo techniques are high-resolution probes of narrow optical homogeneous linewidths \cite{PhysRev.141.391,montjovetbasset2024incoherentmeasurementsub10khz}, exhibiting robustness to laser frequency fluctuations compared to incoherent measurement methods such as spectral holeburning \cite{STAUDT2006720}. Observing a photon echo response requires ensembles of coherent emitters with sufficiently strong collective light-matter interaction characterized by the optical depth (O.D.). Photon echoes also form the basis for ensemble quantum memories such as the atomic frequency comb (AFC) \cite{PhysRevLett.104.040503}, controlled reversible inhomogeneous broadening (CRIB) \cite{https://doi.org/10.1002/lpor.200810056} and revival of silenced echo (ROSE) \cite{Minnegaliev2023}), wherein the optical excitations are coherently stored within an inhomogeneously broadened medium and later rephased in the form of an echo.

We performed photon echo measurements by sending 32~ns long optical pulses with a $\pi$ pulse power of $\sim$ 80~\textmu W, and $\pi/2$ pulse power of $\sim$ 40~\textmu W into the cryostat. The pulse width was kept at the minimum length allowed by the AOM rise/fall times, while the pulse power was optimized to maximize the echo intensity. Fig.~\ref{fig2}(d) shows the photon echo time trace recorded at a temperature of 10~mK and a magnetic field of 200~mT applied in the plane of the chip. The separation of $\tau = 190$~ns between the $\pi/2$ and $\pi$ pulses results in the primary echo appearing at 490~ns and a secondary echo at 680~ns (inset, Fig.~\ref{fig2}(d)). An optically dense medium, enabled by the $\sim$ 0.5~cm long low-loss waveguides, allows us to observe a form of back-action where the primary echo induces an echo emission in the form of the secondary echo \cite{10.1063/5.0176953,PhysRevResearch.2.012026}. The intensities of the $\pi/2$ and $\pi$ pulses relative to the echo in this plot are not to scale due to suppression by the collection AOM and the secondary bump in the pulse and echo time trace is an artifact of reflection between the three excitation AOMs.

Fig.~\ref{fig2}(e) shows the integrated echo amplitudes as a function of the delay $\tau$ between the $\pi/2$ and $\pi$ pulses, for magnetic fields of 0~mT, 120~mT, and 200~mT measured at the base temperature of the sample stage (10~mK). The decay was fitted with the following the expression $\mathrm{I=I_0 exp(-4\tau / T_2)^x}$ with x=1 to account for the observed single-exponential decay \cite{Macfarlane:97,STAUDT2006720}. We observed an increase in optical coherence times $T_2$, from 874~ns at zero magnetic field to 9.7~\textmu s for the highest magnetic field of 200~mT. The highest field coherence time of 9.7~\textmu s corresponds to a homogeneous linewidth of 33~kHz ($\gamma_h= 1/(\pi T_2)$). The observed increase in coherence time with an increasing magnetic field indicates that magnetic noise originating from a paramagnetic environment is the primary cause of dephasing. This noise can arise from several potential sources: impurity spins, spin defects, as well as two-level systems that exhibit magnetic characteristics due to their coupling with rare-earth spins. \cite{PhysRevLett.96.033602,STAUDT2006720}.

Fig.~\ref{fig2}(f) shows the homogeneous linewidth at 200~mT measured as a function of the sample stage temperature. The linewidth follows a linear broadening $\gamma_h = \gamma_0' + \alpha T$ attributed to TLS, reported across amorphous \cite{PhysRevB.94.195138}, polycrystalline \cite{prbsubkhz} and single crystal hosts \cite{klein2025ultralowtemperaturethermodynamicsopticalcoherence}. The low temperature saturation below 150~mK can be attributed to a combination of factors, including the suppression of TLS-phonon scattering at low temperatures \cite{PhysRevB.101.014209}, likely effective sample temperature of $\sim$ 100~mK \cite{prbsubkhz}, and spin pumping effects due to extremely long spin $T_1$. Nevertheless, the model was fit to the data across the entire temperature range for consistency to obtain a broadening rate $\alpha$ of 111~kHz/K. 

% The increase in coherence time with magnetic field at the base temperature of the sample stage (10~mK) as seen in Fig.~\ref{fig2}(e) indicates a paramagnetic spin bath contribution to dephasing. The sources may include impurity spins, spin defects as well as TLS which acquire magnetic character through coupling to rare-earth spins  \cite{PhysRevLett.96.033602,STAUDT2006720}.  

% The observed increase in coherence time with increasing magnetic field in Fig.~\ref{fig2}(e) indicates a paramagnetic spin bath has the dominant contribution to dephasing at the base temperature of the sample stage (10~mK), which can include spin defects and impurities. Furthermore, the TLS can acquire magnetic character through coupling to rare-earth spins \cite{PhysRevLett.96.033602,STAUDT2006720} and may contribute to magnetic field-dependent coherence times.

\begin{figure*}[!th]
    \centering
    \includegraphics[width=0.99\linewidth]{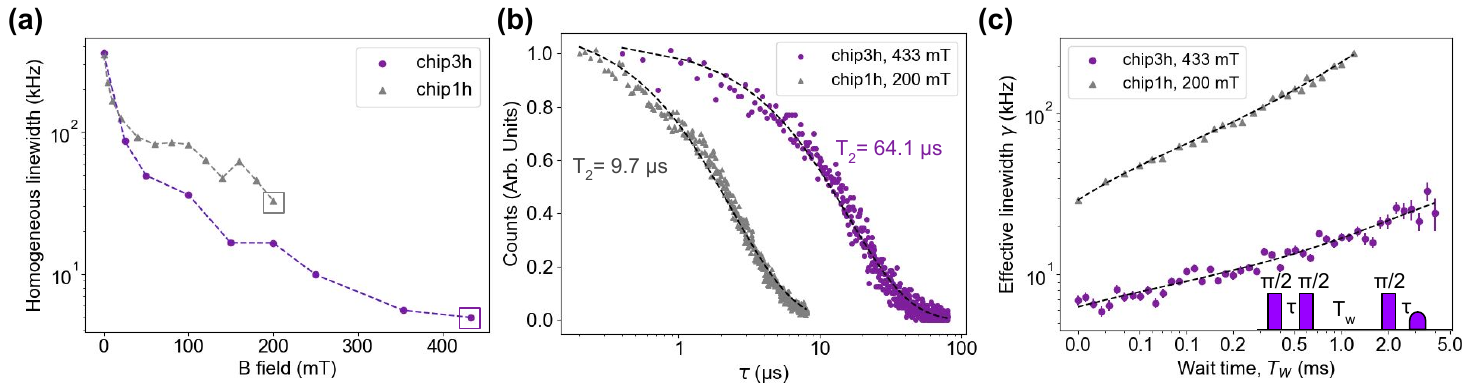}
    \caption{Mitigating optical dephasing and spectral diffusion with annealing and spin-bath polarization. (a) Photon echo homogeneous linewidth versus applied B-field, with gray triangles representing the chip with 1 hour oxygen anneal (chip1h) and purple dots representing an identical chip with 3 hours of oxygen annealing (chip3h). (b) Photon echo decay for the two highest magnetic field points highlighted with a box in (a). (c) Effective linewidth broadening due to spectral diffusion, measured using the three-pulse echo sequence shown in the inset, where the traditional second $\pi$ pulse from a two-photon echo sequence is split into two $\pi/2$ pulses with a variable wait time (T\textsubscript{W}) in between to allow for spectral diffusion.}
    \label{fig3}
\end{figure*}

\subsection{Mitigating optical dephasing and spectral diffusion}

% The lack of strong photon echo modulation (supplementary) suggests a dilute nuclear spin bath in contrast to an abundant nuclear spin bath in well-studied erbium hosts such as YSO\cite{PhysRevApplied.12.024062}, LiNbO\textsubscript{3}\cite{THIEL20101603,doi:10.1126/sciadv.adf4587}.

We analyzed two chips from the same fabrication batch to study the sources of dephasing and spectral diffusion, and to understand the effect of annealing and measurement conditions on mitigating these effects. The chip1h underwent a one-hour oxygen annealing at 400~\textdegree C, while the  chip3h underwent three-hour annealing at the same temperature.

Fig.~\ref{fig3}(a) displays the photon-echo homogeneous linewidth versus magnetic field for the two devices. Although the two samples display a similar homogeneous linewidth of $\sim$ 350~kHz at zero field, the trends diverge with applied magnetic fields. The chip3h chip shows an approximate 2-fold reduction in the homogeneous linewidth for magnetic fields greater than 100~mT. We also investigated the chip3h under magnetic field strengths greater than 200~mT, resulting in a further reduction of homogeneous linewidth down to 5~kHz at a total field magnitude of 433~mT. Fig.~\ref{fig3}(b) displays the corresponding photon echo coherence time: 64.1(9)~\textmu s for chip3h, seven times longer than the 9.7(2)~\textmu s coherence time for chip1h. Paramagnetic defects associated with oxygen deficiency likely dominate dephasing in chip1h, and longer oxygen annealing in chip3h reduced the density of these oxygen vacancies, while higher magnetic fields increased their spin polarization to further suppress the magnetic noise.

In Fig.~\ref{fig3}(c) we study the homogeneous linewidth broadening trend through spectral diffusion to further elucidate the mechanisms of dephasing. We measured spectral diffusion on the two samples using the three-pulse echo sequence \cite{erbiumSDpaper} shown in the inset of Fig.~\ref{fig3}(c). The effective linewidth $\gamma$ due to spectral diffusion in a time scale $T_W$ was measured by sweeping the delay ($\tau$) between the first and second pulses for a fixed waiting time ($T_W$) between the next two and fitting the decay with an exponential. The sequence was repeated for different $T_W$ to observe the linewidth broadening due to spectral diffusion.

Chip1h, measured at 200~mT, displayed a spectral diffusion of 215~kHz occurring within a 1~ms timeframe. Chip3h, when measured at 433~mT, exhibited a reduction in spectral diffusion to 27~kHz over a longer 4~ms timescale. To understand the spectral diffusion mechanisms, data were fitted with a model that includes spectral diffusion due to a paramagnetic spin bath and TLS \cite{dephasingceramicoptical}, given by 
\begin{equation}
    \gamma = \gamma_0 + \gamma_{SD}/2 (1-e^{-RT_W}) + \gamma_{TLS} \log (T_W/t_0),
\end{equation}
where $\gamma_0$ is the linewidth in the absence of spectral diffusion, $\gamma_{SD}$ characterizes the paramagnetic spin bath-induced spectral diffusion, R is the spin bath flip rate, $\gamma_{TLS}$ is the broadening due to the TLS, and $t_0$ is the minimum measurement timescale. 

Table \ref{table1} lists the homogeneous linewidths ($\gamma_h$), along with the fit parameters of the spectral diffusion model for the two samples. The four linewidth parameters ($\gamma_{0} \sim \gamma_h$ , $\gamma_{SD}$, $\gamma_{TLS}$) all show a several-fold reduction in the sample subjected to additional oxygen annealing. Thus, dephasing and spectral diffusion were significantly reduced with longer annealing and higher magnetic fields.

Interestingly, we note that $\gamma_{SD} >> $ $\gamma_{TLS}$ for both datasets, indicating that the paramagnetic spin bath is the dominant source of spectral diffusion over TLS in both samples. Comparing the relative fit of the TLS-only and spin-bath-only models (Supplemental Material \cite{suppl}) further supports this hypothesis. This is in contrast to TLS-dominated spectral diffusion observed in Er-doped Y\textsubscript{2}O\textsubscript{3} nanoparticles at 3~K \cite{PhysRevB.108.075107} or Er-doped silica fiber at sub-Kelvin temperatures \cite{PhysRevB.94.195138}. This indicates an effective freezing of TLS modes in these Er-doped \TiOtwo\ films at the measurement conditions considered here.

Consequently, paramagnetic defects are the main drivers of optical dephasing and spectral diffusion in both samples. These defects are shown to be suppressed through the application of magnetic fields at sufficiently low temperatures and extended oxygen annealing, indicating that they have their origin in oxygen deficiency of the thin films. While we clearly detected signatures of TLS in both chips, they did not significantly influence spectral diffusion or the homogeneous linewidth.

We expect the narrowest homogeneous linewidth of 5~kHz to be limited by charge noise and a residual unpolarized paramagnetic spin bath \cite{FOSSATI2023120050}. The inversion symmetry at the titanium site in the rutile phase of \TiOtwo\ could lead to further improved coherence properties over the anatase phase studied here \cite{10.1063/5.0222269}. Homogeneous linewidth measurements in Fig.~\ref{fig3}(a) when repeated at lower optical powers did not show any power dependence (Supplemental Material \cite{suppl}), ruling out instantaneous spectral diffusion (ISD) \cite{PhysRevB.41.6193,erbiumSDpaper} due to Er-Er interactions as a significant source of decoherence.

\begin{table}[h]
\begin{tabular}{|l|c|c|}
\hline
 & \bfseries chip1h at 200~mT  & \bfseries chip3h at 433~mT  \\
 \hline
$\gamma_h$ (kHz) & 32.8 (0.5) & 5.0 (0.1) \\ \hline
$\gamma_0$ (kHz) & 26.3 (3.2)  &  6.2 (0.7)  \\ \hline
$\gamma_{SD}$ (kHz) & 1347 (1625)  &  42.8 (17.8) \\ \hline
$\gamma_{TLS}$ (kHz) & 15.8 (3.8) &  1.4 (0.6) \\ \hline
R (kHz) & 0.27 (0.3) & 0.3 (0.2) \\ \hline
\end{tabular}
\caption{Homogeneous linewidth and spectral diffusion fit parameters for the 1 hour annealed chip and the chip with 3~hour annealing.}
\label{table1}
\end{table}

\subsection{Spin relaxation times}

\begin{figure}[!t]
    \centering
    \includegraphics[width=0.8\linewidth]{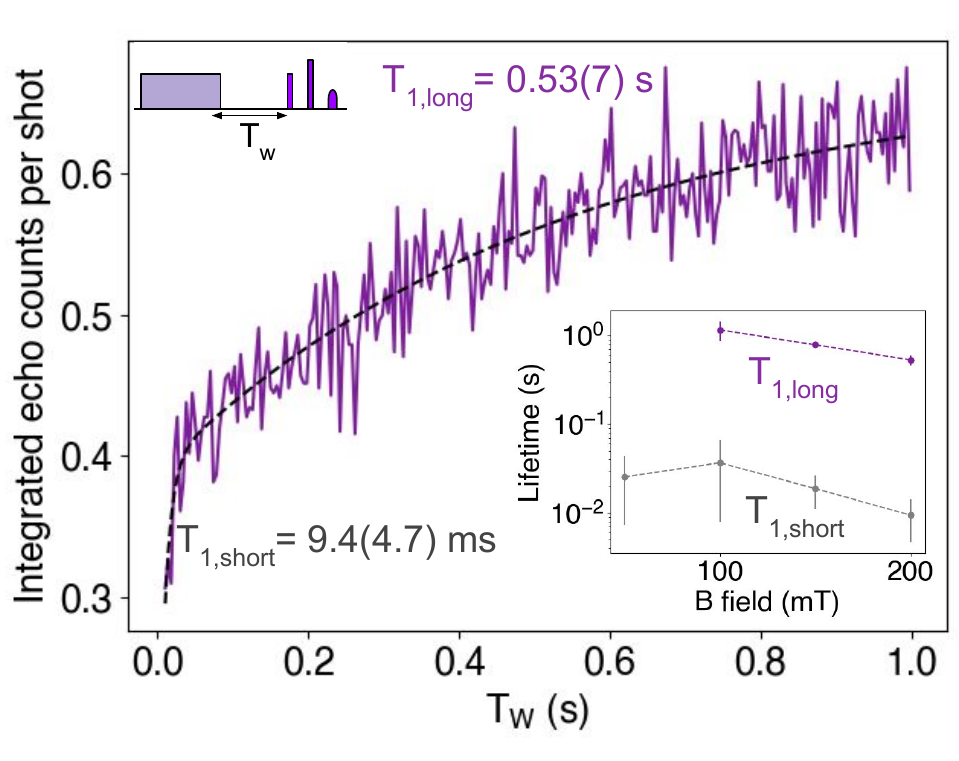}
    \caption{ Electron spin relaxation time measured with an all-optical technique. (a) An example of spin lifetime for chip1h measured at 200~mT measured using the photon echo saturation pulse sequence shown in the upper left inset. Top left inset, an optical excitation pulse (1~ms) is used to shelve the population in the ground state Zeeman sub levels, and a photon echo sequence is used to probe relaxation after a wait time, $T_W$. The resultant echo amplitude recovers on two timescales characterized by the time constants ${T_{1,short}}$ ($\sim$ 10~ms) and ${T_{1,long}}$ $(\sim$ 1~s). Bottom right inset, is a plot of the two population relaxation time constants, ${T_{1,short}}$ and ${T_{1,long}}$ versus applied B-field. }
    \label{fig4}
\end{figure}
The optical coherence time of Er ions is limited by the optical lifetime on the order of milliseconds, and storage for longer durations requires long-lived electron spin sublevels. The inset of Fig.~\ref{fig4} shows the sequence used to measure the electron spin lifetime with all-optical techniques. A 1~ms-long optical saturation pulse was used to excite the Er population from the ground state $\mathrm{^4I_{15/2}}$ to the optically excited state $\mathrm{^4I_{13/2}}$. As illustrated in the inset of Fig.~\ref{fig1}(d), when the spin lifetime is significantly longer than the optical lifetime, a fraction of the population will be shelved and show recovery on a timescale determined by the spin $T_1$, also called persistent holeburning.  A photon echo sequence is sent with a waiting time ($T_W$) between 10~ms and 2~s after the saturation pulse as a probe of the ground state population, and the integrated echo amplitude shows recovery on timescales commensurate with the population relaxation times.

Fig.~\ref{fig4} shows the population relaxation measured on chip1h at 200~mT, with a short relaxation time component ${T_{1,short}}$ of 9.4(4.7)~ms and a long relaxation time component ${T_{1,long}}$ of 0.53(7)~s. The lower right inset of Fig.~\ref{fig4} shows the two relaxation time constants plotted as a function of the magnetic field. We observe an increase in ${T_{1,long}}$ with decreasing magnetic fields, which suggests direct process-limited electron spin relaxation as the population relaxation mechanism \cite{Raha2020}. We note that disorder in the film may result in magnetic field scaling of spin $T_1$ between $\sim B^{-1.x}$ for amorphous systems \cite{PhysRevB.92.241111,bornadel2024holeburningexperimentsmodeling} and $B^{-4}$ for single crystals \cite{Raha2020}. The longest $T_1$ of 1.1~s was measured at 100~mT, comparable to Er ions in bulk crystals \cite{Raha2020}. Measurements could not be performed for magnetic fields below 100~mT due to spin pumping effects from the photon echo probe sequence, likely because $T_1$ is much longer than 1~s.  The lack of strong spin-spin relaxation may be attributed to broad inhomogeneous linewidths effectively suppressing Er-Er interactions \cite{PhysRevApplied.19.044029}. While there may be subclasses of ions with faster spin relaxation \cite{bornadel2024holeburningexperimentsmodeling,PhysRevApplied.19.044029}, the short component with a timescale of $\sim$ 10~ms did not show significant variation with the magnetic field, suggesting that this recovery can be attributed to residual optical relaxation instead of spin relaxation. Although we did not repeat magnetic field dependent lifetime measurements on chip3h, a comparable lifetime of 1.63~(0.91)~s was measured at $B=200$~mT on the chip (Supplemental Material \cite{suppl}).

\subsection{Stark control of Er emission}
\begin{figure}[!t]
    \centering
    \includegraphics[width=0.7\linewidth]{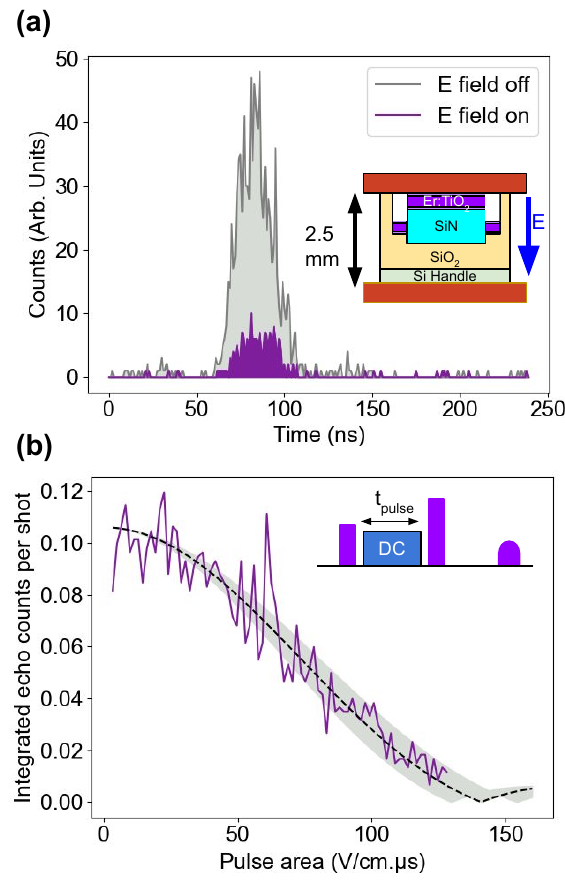}
    \caption{ Stark modulation of the Er echo emission (a) Echo envelope with and without applied electric field pulse showing suppression, inset shows the device configuration (b) Integrated echo amplitude modulation with varying Stark (electric field) pulse area. Black dashes show the model with gray band corresponding to  10 \% variation in the estimated Stark coefficient, inset shows the pulse sequence.}
    \label{fig5}
\end{figure}
Electrical control of Er emission through the linear Stark effect is necessary for on-demand quantum memory protocols and tunable single photon emitters. We perform Stark modulation of the echo envelope as a method to demonstrate the electrical control of Er emission on the platform. An electric field is applied perpendicular to the plane of the chip using electrodes integrated on the top and bottom of the chip, separated by $\sim$ 2.5~mm (Fig.~\ref{fig5}(a), inset). Fig.~\ref{fig5}(a) shows the echo envelope corresponding to no Stark pulse and the longest Stark pulse length of 3~\textmu s. We achieve 90\% extinction of the echo envelope for the longest Stark pulse width of 3~\textmu s, limited by the maximum separation $\tau$ = 3~\textmu s between the $\pi$/2 and $\pi$ pulses. The maximum Stark pulse area of 120~V/(cm$\cdot$\textmu s) was limited by the setup; in particular, given the electrode separation of 2.5~mm, the average E-field magnitude was limited to 40~V/cm. Further, the orthogonal orientation of the DC E-field and the light polarization in the waveguide resulted in the ions with the weakest DC Stark shift contributing the strongest to the echo.

The dependence of echo amplitude on the Stark pulse area can be used to estimate the Stark coefficient of Er\cite{doi:10.1021/acs.nanolett.0c02200}. The bottom inset of Fig.~\ref{fig5}(b) shows the pulse sequence with a voltage pulse of a variable length $t_{pulse}$ added between the $\pi/2$ and $\pi$ pulses of a standard photon echo, separated by 3 \textmu s. Fig.~\ref{fig5}(b) shows the integrated echo area as a function of the Stark pulse area and the black dashed lines indicate the numerically modeled echo response corresponding to an estimated Stark sensitivity k = 5.8~kHz/V/cm with the gray band indicating the uncertainty of $\pm$~0.5~kHz/V/cm. The details of the model are discussed in Supplemental Material \cite{suppl}. We note that the estimated Stark coefficient is 3 times lower than the values reported for Er in other hosts such as LiNbO\textsubscript{3} (18-25~kHz/V/cm) \cite{PhysRevLett.131.170801,HASTINGSSIMON2006716} and Y\textsubscript{2}SiO\textsubscript{5} (20~$\pm$~5.8~kHz/V/cm) \cite{cpl_40_7_070301}, which may be attributed to a non-polar symmetry of \TiOtwo .

\begin{table*}[!t]
\begin{tabular}{|l|l|l|}
\hline
 &  {\bf Single ion quantum memories} & {\bf Ensemble quantum memories} \\
 \hline 
Memory Protocol & Emissive (Barrett-Kok) \cite{PhysRevA.71.060310}   & Absorptive (AFC \cite{PhysRevResearch.3.023099}, ROSE/SEMM \cite{doi:10.1021/acs.nanolett.0c02200,Arcangeli2016StarkEM})  \\ \hline
Storage time  &  High (Spin T\textsubscript{2},  1~ms - 1~s) (projected) &  Medium (Optical T\textsubscript{2}, $\sim$ 64~\textmu s) \\ \hline
Bandwidth & Narrow  (1/Optical T\textsubscript{1} $\sim$ 10~kHz)  & High, up to 36~GHz ($\Gamma_{\mathrm{Inh}}$)\\ \hline
% Multiplexing Capacity & Medium (Spectral and channel) &  High (Spectral, temporal and channel) \\ \hline
Device geometry & Nanophotonic cavity & cm-long waveguide or nanophotonic cavity \\ \hline
% Challenge & Photon indistinguishability & Maintaining optical T\textsubscript{2} with high-optical depth \\ \hline

\end{tabular}
\caption{Quantum memory metrics for back-end deposited Er:\TiOtwo\ films on foundry photonics. Numbers based on current performance in the platform.}
\label{table2}
\end{table*}

\section{On-chip quantum memories based on Er:\TiOtwo\ on foundry photonics}
Having established that this BEOL-deposition on foundry photonics platform fulfills the requirements for a quantum memory, this section evaluates prospects of ensemble and single-ion-based devices. As summarized in Table \ref{table2}, these two memory modalities operate in complementary regimes. Ensemble memories function in an absorptive regime, offering the advantages of broadband and multimode storage; however, their storage duration is limited by optical coherence times in the absence of spin-wave transfer. In contrast, single-ion memories function emissively, featuring an intrinsic spin-photon interface and inherent heralding. This enables substantially longer storage times, limited only by spin coherence, but at the cost of a narrower optical bandwidth and reduced multimode capacity compared to ensembles.

\subsection{Single ion-based quantum memories}

Single-ion quantum memories intrinsically support heralding and quantum logic operations, which are foundational for networking schemes that rely on quantum error correction \cite{Muralidharan2016}. However, the weak optical dipole moments of rare-earth ions necessitate photonic Purcell enhancement for addressing individual ions. Small mode volume and high quality factor silicon photonic crystal cavities have enabled the readout of single Er ions in backend deposited \TiOtwo\ films by achieving Purcell factors of $\sim$ 500 \cite{pettit2025monolithicallyintegratedcbandquantum}.

Achieving photonic indistinguishability, a fundamental benchmark for single-emitter quantum memories \cite{PhysRevA.71.060310,PhysRevX.15.011071}, necessitates transform-limited homogeneous linewidths and slow spectral diffusion \cite{Kambs_2018}. The homogeneous linewidth of 5~kHz (Fig.~\ref{fig3}(a)) and spectral diffusion of 27~kHz over 4~ms (Fig.~\ref{fig3}(c)) measured in this work are both comparable to the Purcell-broadened radiative linewidth of $1/(2\pi\: T_1) = 11.5$~kHz of single Er ions in dilute Er-doped films via BEOL-deposition on foundry-fabricated silicon nanocavities  \cite{pettit2025monolithicallyintegratedcbandquantum}. Although the single Er ion linewidth of $\sim$ 10~MHz at 4~K \cite{pettit2025monolithicallyintegratedcbandquantum} is substantially broader than the 5~kHz photon echo homogeneous linewidth at 10~mK, the latter broadens to a comparable value of 1~MHz when measured at 3.8~K (Supplemental Material \cite{suppl}). The remaining discrepancy is likely due to a combination of spectral diffusion and laser linewidth broadening over the long integration times needed for single-ion spectral scans ($\sim$ seconds). Nevertheless, the photon echo measurements shown here provide a reliable proxy for the platform's intrinsic homogeneous linewidths at sub-Kelvin temperatures, and a similar reduction in the single-ion linewidth is anticipated under optimized measurement and annealing conditions. These findings indicate that photon indistinguishability in this platform can be realized in a similar manner to bulk Er-based nanophotonic platforms \cite{Ourari2023}.

A spin coherence time of at least a millisecond is required for entanglement distribution over a metropolitan area quantum network \cite{PhysRevApplied.19.044029}. While spin coherence times were not measured in this work, the optical coherence time of 64~\textmu s suggests a low magnetic noise environment, and we expect Er electron spin coherence times exceeding optical coherence times \cite{Ortu2018}, limited only by the measured spin lifetime of 1~s. The presence of a weak titanium nuclear spin bath is manifested as a modulation in the photon echo signal \cite{PhysRevB.102.115119} (Supplemental Material \cite{suppl}), indicating its potential as a long-lived quantum memory register \cite{PRXQuantum.4.010323,Ruskuc2022}.

While the DC Stark control was demonstrated on an ensemble of ions in the anatase phase (Fig.~\ref{fig5}), this technique is also applicable for spectrally matching individual Er ions to facilitate Bell state measurements \cite{PhysRevLett.131.170801, cpl_40_7_070301}.

\subsection{Ensemble-based quantum memories}

Observing collective photon echo response signifies that the platform is suited for all-optical ensemble quantum memories. Quantum memories based on atomic frequency combs (AFC, Stark modulated-AFC) \cite{PhysRevResearch.3.023099} and echo-silencing based protocols, such as Stark echo modulation memory (SEMM), \cite{doi:10.1021/acs.nanolett.0c02200,Arcangeli2016StarkEM} enable broadband, on-demand, multimode storage of photonic qubits. Storage times up to $\sim$ 64~\textmu s are possible, limited only by the current Er optical coherence times in our system. While electron spin-wave storage would be challenging in polycrystalline films due to an anisotropic g tensor of Er, zero-field hyperfine transitions in \textsuperscript{167}Er could be used for longer storage times up to milliseconds \cite{eropticallyexcitedstate,Rajh_2022}. 

The AFC protocol is attractive due to an inherently low-noise storage; however, its storage efficiency depends on burning several high-extinction, narrow persistent holes in the optical absorption spectrum. Persistent holeburning has traditionally been difficult to achieve in Er-doped single crystals because of strong spin-spin interactions \cite{Askarani:20,Saglamyurek2015}. Recent works have shown that a combination of broad inhomogeneous linewidths and sub-Kelvin temperatures can suppress  spin-spin interactions \cite{bornadel2024holeburningexperimentsmodeling}. Spin lifetimes of approximately 1 second in our system (Fig.~\ref{fig4}) are significantly longer than the optical lifetimes of 2.8 ms, suggesting that efficient holeburning is indeed possible in our platform. Long-lived hyperfine transitions of \textsuperscript{167}Er \cite{Asadi2020,spinechogroundandopter,WANG2023119935} could be alternatively used for population shelving at zero magnetic fields \cite{Yasui:22,Yasui:24}. 

% Both SEMM and SM-AFC protocols enable on-demand, electrically-controlled retrieval from the memory. However, a fast Stark pulse of length significantly shorter than the optical coherence times is required to perform  complete echo extinction and a subsequent on-demand recall. We have achieved 90 \% extinction of the echo amplitude with a 3 \textmu s long Stark pulse in this work, which was limited by the 2.5 mm electrode spacing as shown in Fig.~5(a). By integrating on-chip electrodes with a tighter 0.2 mm spacing, we anticipate achieving complete echo extinction with Stark pulses as fast as 95 ns. This improvement will allow for on-demand memory recall within the 64 \textmu  s optical coherence time (supplementary section S6).

Quantum memory protocols based on echo silencing like SEMM do not require complex preparation steps like the AFC; instead, they rely on efficient suppression of the noisy primary echo and subsequent emission of a noise-free secondary echo. We achieved 90\% suppression of the primary echo with a 3~\textmu s long Stark pulse in Fig.~\ref{fig5}(a), constrained by the 2.5~mm wide electrode spacing. By implementing on-chip electrodes with a reduced spacing of 0.2~mm, the required Stark pulse duration for full extinction is expected to decrease to 95~ns. This enhancement would enable the implementation of an SEMM memory with complete echo extinction and recall within the 64 \textmu s optical coherence time (Supplemental Material \cite{suppl}). Additionally, our photonic substrate-agnostic integration strategy allows us to leverage the strong electro-optical linearity of platforms such as lithium niobate to implement echo-silencing based on rapidly tunable nanocavities \cite{PhysRevA.108.012614}. Doping erbium directly in lithium niobate \cite{PhysRevApplied.18.014069}  also offers the advantages of electro-optical tunability, however, this benefit is offset by the high nuclear spin abundance in LiNbO\textsubscript{3} where for both Li and Nb have no stable isotopes with I=0, in contrast to \TiOtwo\ where Ti has 87\% abundance of I=0.

 A device optical depth of at least 1 is necessary for high-efficiency optical quantum memories. The silicon nitride waveguide devices used in this work have an estimated Er optical depth $\sim$ 0.015. An optical depth of 1 is projected to be achievable through a combination of enhancements, such as increasing the Er concentration, depositing a thicker doped \TiOtwo\ layer, and increasing the low-loss SiN waveguide length to centimeter-scale or transitioning to impedance-matched Si nanophotonic cavities \cite{PhysRevA.82.022310,PhysRevApplied.12.024062} (Supplemental Material \cite{suppl}), making high-efficiency quantum memories feasible on the platform.

% With an anatase TiO\textsubscript{2} grain size of approximately 13.5(3) nm, we could potentially leverage the modified phonon density of states to further suppress spin-lattice relaxation and enable efficient spectral holeburning \cite{modifcationofphonon}.  The high densities of rare-earth ions (50-100 ppm) required to reach optical depth $\sim$ 1 are prohibitive for Kramers REIs like erbium due to the resultant strong electron spin-spin interaction. Thus maintaining long optical coherence and spin relaxation times while simultaneously maximizing optical depth are the primary challenges with erbium ensemble quantum memories. Long waveguides or impedance-matched optical cavities have been proposed to maximize the collective light-matter interaction with some macroscopic quantum memory demonstrations on bulk crystals. The limited photonic integration on the state-of-the-art bulk crystals has been the primary challenge towards high-efficiency scalable, photonic integrated quantum memories. We analyze our platform with respect to the key requirements for ensemble quantum memories.

\section{Conclusion}

We studied Er-doped \TiOtwo\ thin films integrated with foundry-fabricated nanophotonics via backend deposition as a scalable photonic quantum memory platform. We performed high-resolution photon echo spectroscopy to study Er optical dephasing mechanisms at sub-Kelvin temperatures, which enabled us to suppress dephasing through optimized measurement conditions and extended annealing. This yielded kHz-level homogeneous linewidths and spectral diffusion, competitive with state-of-the-art for Er in nanophotonic devices \cite{Weiss:21,RinnerBurgerGritschSchmittReiserer+2023+3455+3462,Ourari2023,doi:10.1126/sciadv.adu0976,PhysRevX.12.041009}. We show electrical control of Er emission through the DC Stark effect and second-long spin lifetimes in the platform, making Er:\TiOtwo\ \cite{10.1063/5.0222269} suitable for both single-ion-based emissive memories and ensemble-based absorptive quantum memories.

% Our thin-film back-end integration strategy is not restricted to the choice of specific rare-earth ions and host materials or photonic substrates. Disorder arising from film growth on structured photonic substrates can be addressed through post-processing and optimized measurement conditions. 

The versatility of the backend integration strategy is demonstrated by its successful implementation on diverse foundry photonic platforms, including the silicon nitride waveguides of this study and previously reported silicon nanocavities \cite{pettit2025monolithicallyintegratedcbandquantum}. This approach will enable the monolithic integration of quantum memories with other photonic platforms that are uniquely suited for on-chip quantum light generation, linear and nonlinear photonic elements, or detection capabilities \cite{PRXQuantum.5.010101}.

% . Such integration would unlock on-chip capabilities for generating, storing, and routing quantum light, for applications in long-distance communication \cite{Kimble2008}, synchronized single-photon generation \cite{PhysRevLett.131.033601}, and asynchronous Bell state analysis \cite{nakav2025quantumcnotgateactively,doi:10.1126/sciadv.1501772}. 

\begin{acknowledgments}
Authors  S. G., R.M.P., A.S., S.D.J., M.K.S., and S.E.S. acknowledge support from the U.S. Department of Energy, Office of Science, Advanced Scientific Computing Research (ASCR) program under Grant No. CRADA A22112 through the Chain Reaction Innovations program at Argonne National Laboratory. V.N. and A.M.D. were supported by the U.S. Department of
Energy, Office of Science, Advanced Scientific Computing Research (ASCR)
program under contract number DE-AC02-06CH11357 as part of the InterQnet quantum networking project.
\end{acknowledgments}

% \section*{Data Availability Statement}

% \appendix

% \section{Appendixes}

\nocite{}
\bibliography{references}% Produces the bibliography via BibTeX.

\newpage

\title{Erbium Quantum Memory Platform with Long Optical Coherence via Back-End of Line Deposition on Foundry-Fabricated Photonics: Supplemental Material}

% \title{ High Optical Coherence in Erbium Quantum Memory Platform Integrated via Backend Deposition with Foundry Photonic Circuits: Supplemental Material}

\maketitle

\makeatletter
\renewcommand \thesection{S\@arabic\c@section}
\renewcommand\thetable{S\@arabic\c@table}
\renewcommand \thefigure{S\@arabic\c@figure}
\makeatother
\setcounter{equation}{0}
\setcounter{figure}{0}
\setcounter{table}{0}
\setcounter{page}{1}
\setcounter{section}{0}

\section{Material growth and Device Packaging}
\TiOtwo\ films were grown using molecular beam deposition from titanium tetraisopropoxide (TTIP) precursors with growth described in previous work \cite{10.1063/5.0224010}. Two identical chips were oxygen-annealed in a tube furnace at temperatures of 400~\textdegree C. One of the chips was annealed for 1 hour (chip1h) while another chip was annealed for 3 hours (chip3h). Followed by film growth and annealing, both chips were packaged with lensed fibers at PLC Connections. For the first device (chip1h), the fiber-to-fiber device transmission was measured to be $\sim$ 13~dB at 1520~nm at cryogenic temperatures ($\sim$4~K), the second device (chip3h), exhibited a lower device transmission loss of $\sim$ 8.5~dB at 1532~nm.  The difference in the device transmission between the two devices may be attributed to the packaging process and the degradation in chip1h transmission from 3 thermal cycles.

We estimated a \TiOtwo\ refractive index of 2.475 on chip1h using the 2.15 nm FSR of the MZI interference pattern. Deposition of 60 nm \TiOtwo\ did not result in an increase in the silicon nitride waveguide losses of $\sim$ 2.5 dB/cm. The $\sim$ 9.9 mm long waveguide used in the measurement consisted of a 9.4 mm long segment clad in oxide and a 0.486 mm long segment exposed in MZI arm where \TiOtwo\ was deposited. 

% \begin{figure}[!th]
%     \centering
%     \includegraphics[width=0.99\linewidth]{V2 PR Applied Format/SI/S1.pdf}
%     \caption{Thin-film integration workflow on foundry photonics chips}
%     \label{figs1}
% \end{figure}

\section{Experimental Setup}
Measurements on the reference chip used MOGlabs Tunable Cateye SAF laser, while a Toptica
CTL1550 was used for the measurements on the chip3h in Fig.~3 of the main text due to its easier tunability. We do not expect the photon echo linewidths here to be limited by the laser linewidths. The AOM used for pumping and collection was Aerodiode 1550AOM-2, variable optical attenuators (VOA) were Thorlabs V1550A. Superconducting nanowire single photon detectors (SNSPD) included systems from Single Quantum and Quantum Opus. Pulse generation and time-tagging was handled by Quantum machines OPX. A magnetic field was applied using a 3 axis vector magnet (AMI Model 430) with (0.25-0.25-0.25)~T magnetic fields. The magnetic field was applied in the plane of the chip, using the magnetic X axis across this manuscript, except for the measurement in Fig.~3 on the chip with higher annealing, where $B > 0.25$~T was applied by turning on the Y and Z axis up to 0.25~T.

\section{Spectral diffusion modeling}

\begin{figure}[!th]
    \centering
    \includegraphics[width=0.99\linewidth]{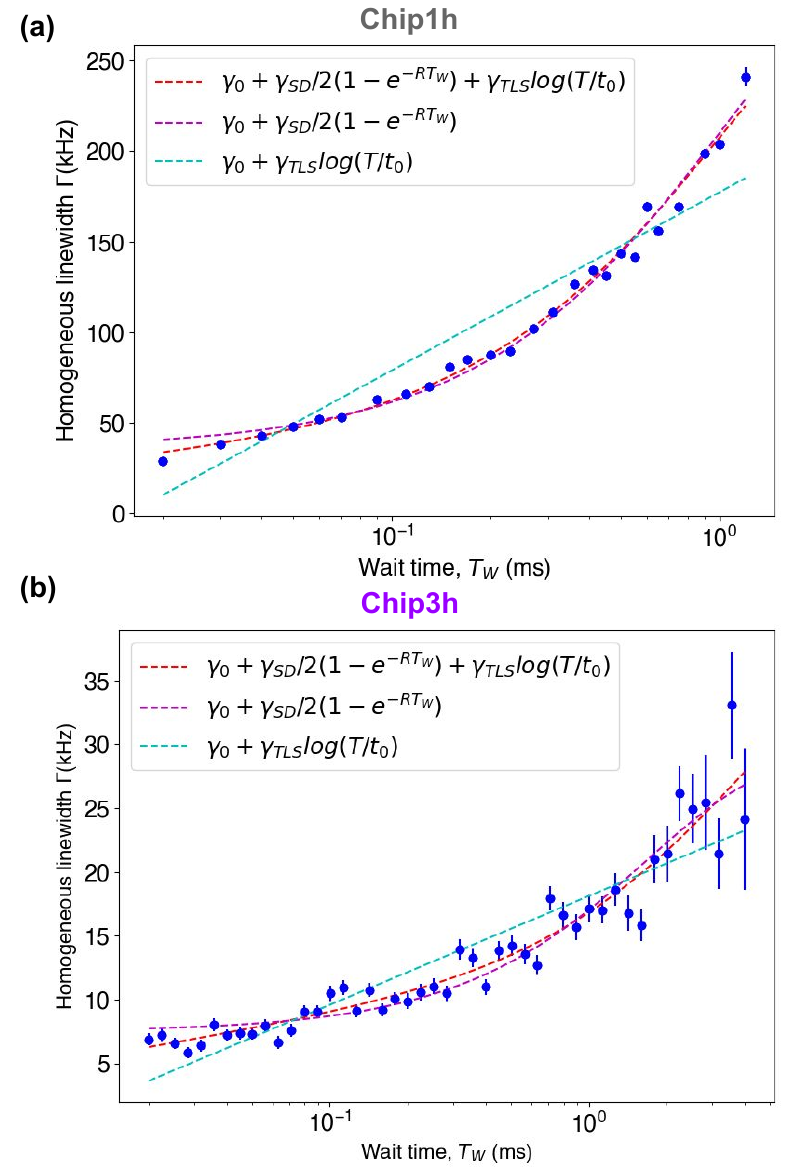}
    \caption{Modeling spectral diffusion for (a) Chip1h, 1 hour $\mathrm{O_2}$ anneal (b) Chip3h, 3 hour $\mathrm{O_2}$ anneal.}
    \label{figs2}
\end{figure}

Fig.~\ref{figs2} below compares the spectral diffusion model fitted to the data on the chip1h Fig.~\ref{figs2}(a) and the chip3h Fig.~\ref{figs2}(b). We compare three models:

\begin{itemize}
    \item  TLS only: The TLS only model shown with cyan dashes diverges with both the datasets in Fig.~\ref{figs2}(a) and Fig.~\ref{figs2}(b).
    
    $\gamma= \gamma_0 + \gamma_{TLS} log(T/t_0)$
    \item  Paramagnetic spin bath only: While the paramagnetic spectral diffusion only model shown in magenta dashes shows closer agreement over the TLS only model, it shows some divergence in the low $T_W$ and high $T_W$ limit.
    
    $\gamma= \gamma_0 + \gamma_{SD}/2(1-e^{-RT_W}) $
    \item  Paramagnetic spin-bath + TLS: The full model which includes paramagnetic and TLS induced spectral diffusion shown with red dashes fits both the datasets across all values of $T_W$. 
    
    $\gamma= \gamma_0 + \gamma_{SD}/2(1-e^{-RT_W}) + \gamma_{TLS} log(T/t_0)$
\end{itemize}

\section{Power dependence of homogeneous linewidth}

\begin{figure}[!th]
    \centering
    \includegraphics[width=0.99\linewidth]{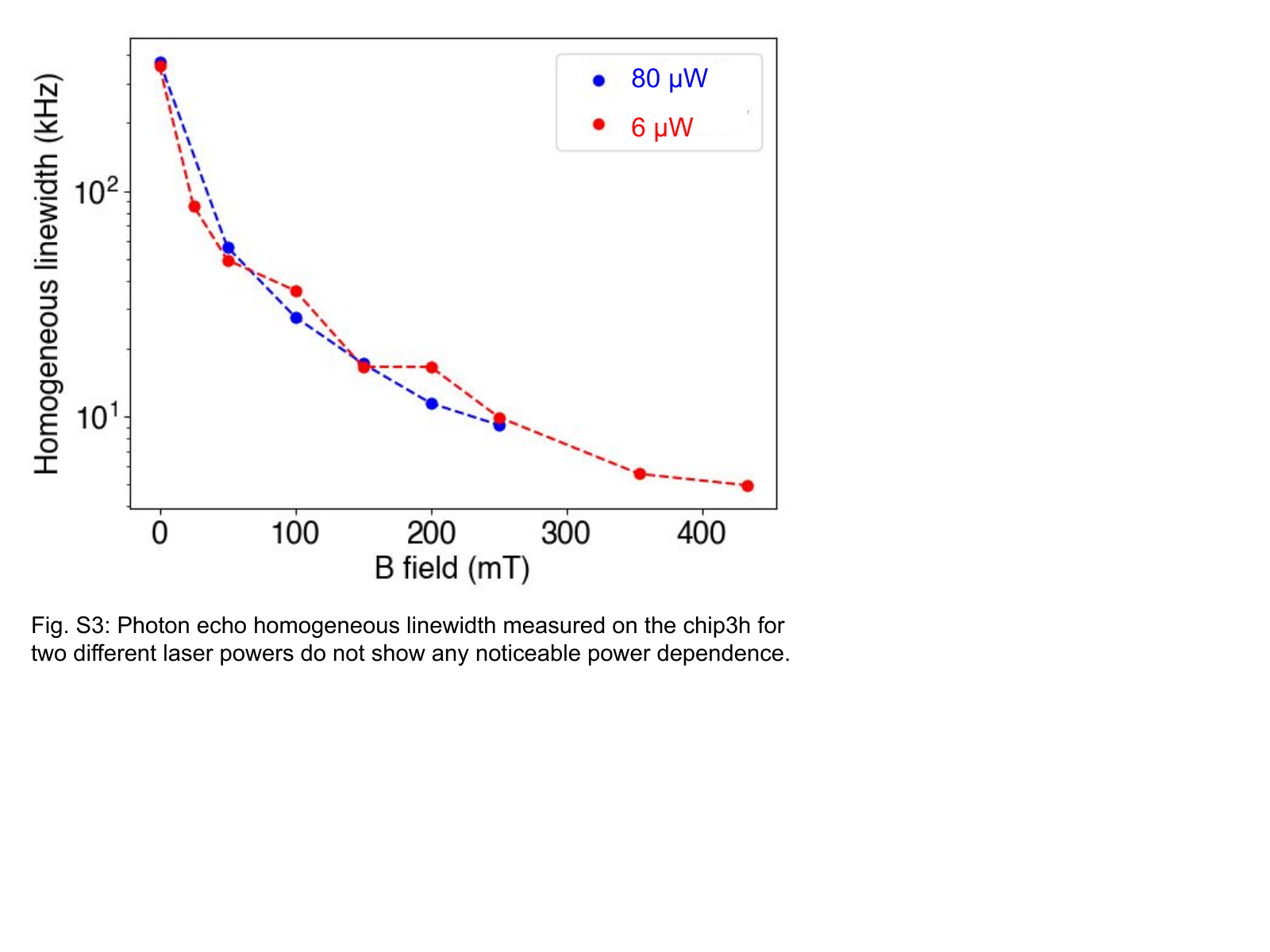}
    \caption{Photon echo homogeneous linewidth on chip3h for two difference powers repeated with magnetic fields shows no power dependence}
    \label{figs3}
\end{figure}

Fig.~\ref{figs3} displays the photon echo homogeneous linewidth at $\pi$ pulse laser power of 80~\textmu W (red) and 6~\textmu W (blue). The $\pi/2$ pulse power was scaled to 0.5 times the $\pi$ pulse power for each measurement. These measurements were conducted on the chip annealed for 3 hours, with the 80~\textmu W data (blue dots) previously presented in Figure 3(a) of the main text. The consistency between the two linewidth trends across various magnetic fields suggests that coherence times exhibit no significant dependence on laser power for the measurement powers in the main text. 

\section{Spin lifetime in the chip3h sample}
Fig.~\ref{figs4} below shows the spin lifetime of 1.63 s measured for the 3-hour annealed sample at a magnetic field of 200~mT. A short component of the decay with $\sim$ 10~ms lifetime, while not observed due to a low signal-to-noise ratio and a coarser temporal sampling than the main text, could not be completely ruled out. 

\begin{figure}[!th]
    \centering
    \includegraphics[width=0.99\linewidth]{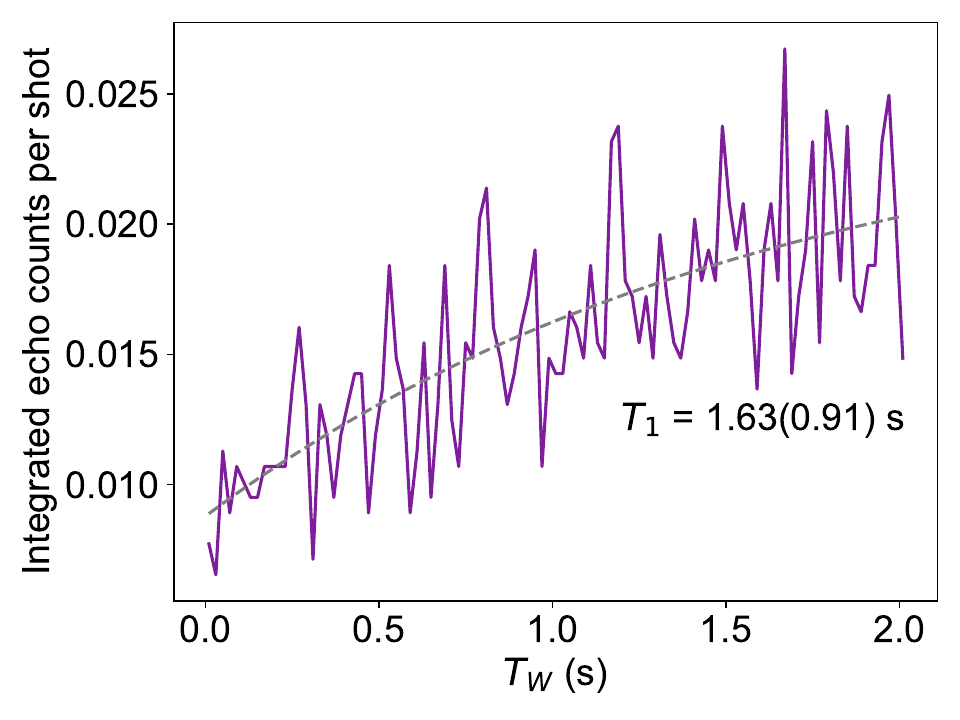}
    \caption{Spin lifetime measured on chip3h (3-hour O\textsubscript{2} anneal for $B=200$~mT.}
    \label{figs4}
\end{figure}

\section{Stark echo modulation modeling}
\begin{figure}[!th]
    \centering
    \includegraphics[width=0.99\linewidth]{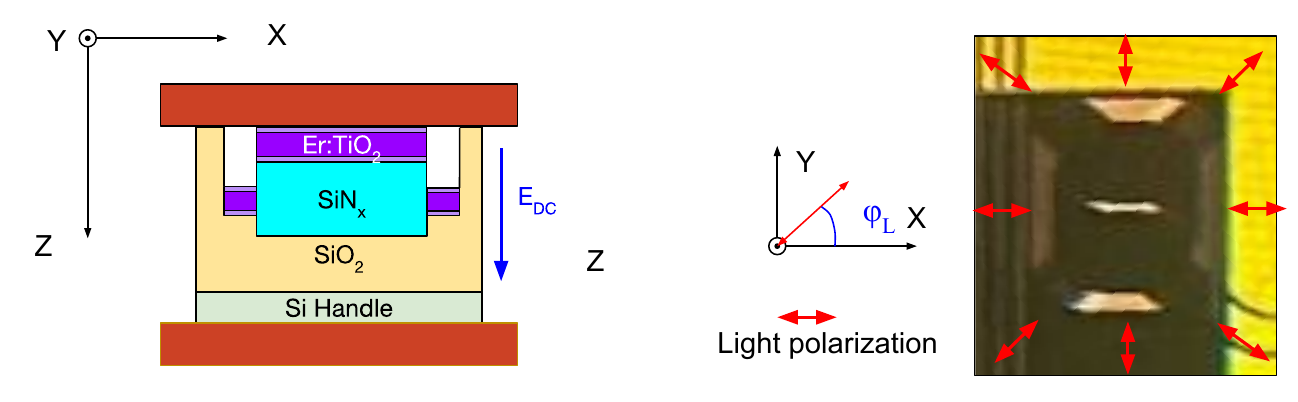}
    \caption{Orientation of the DC Electric field (blue arrow) and the light polarization axis (double sided red arrow) with respect to the cartesian coordinates used for modeling the Stark echo response.}
    \label{figs5}
\end{figure}

We follow the expression given by equation S1 in Ref. \cite{doi:10.1021/acs.nanolett.0c02200}, where the echo amplitude was derived for an ensemble of emitters with a symmetric distribution of linear Stark shifts, in systems such as in polycrystalline thin films, nanoparticles, or ceramics. Assuming a configuration of orthogonal DC E-field and the AC E-field of the waveguide mode, the echo amplitude can be modeled as \cite{doi:10.1021/acs.nanolett.0c02200}
\begin{equation}
\begin{aligned}
    A = A_0 \int &C(\theta, \theta_L, \phi, \phi_L) \cos (2 \pi k E t_{pulse} \cos \theta) \times \\ &\sin \theta \sin \theta_L d\theta d\theta_L d\phi d\phi_L
\end{aligned}
\end{equation}
where the DC E field is along the Z axis, light propagation is in the XY plane and the light polarization in the waveguide rotates in the range $\phi=\left[0, \pi\right]$ due to the spiral geometry of the waveguide (Fig.~\ref{figs5}). $E$ is the DC electric field magnitude, $t_{pulse}$
is the Stark pulse length and k is the Stark coefficient for Er in the Anatase phase. The integral is over $\theta$, where $\theta$ is the angle between the Er’s electric dipole moment orientation in the Anatase site and the Z-axis.

Since light polarization is along the XY plane (Fig.~\ref{figs5}), $\theta_L=\pi/2$ and the term $C(\theta, \theta_L, \phi, \phi_L)$ from eq.~S2 in ref\cite{doi:10.1021/acs.nanolett.0c02200} is defined as:
\begin{equation}
\begin{aligned}
    C(\theta, \theta_L, \phi, \phi_L) = | &\sin \theta  \cos \phi  \sin \theta_L  \cos \phi_L + \\
    &\sin \theta \sin \phi  \sin \theta_L \sin \phi_L + \cos \theta \cos \theta_L|
\end{aligned}
\end{equation}
\noindent can be reduced to:
\begin{equation}
    C(\theta, \theta_L, \phi, \phi_L) = | \sin \theta ( \cos \phi  \cos \phi_L +
    \sin \phi \sin \phi_L)  |
\end{equation}

Substituting for $ C(\theta, \theta_L, \phi, \phi_L)$, the integral over $\theta_L, \phi, \phi_L$
is trivial and adds an overall constant to the term $A_0$, with the resultant expression being identical to eq.~S3 in ref\cite{doi:10.1021/acs.nanolett.0c02200} we used in the fit in Fig.~5 of the main text
\begin{equation}
    A = A_0' \int_0^{\pi/2}  \cos (2 \pi k E t_{pulse} \cos \theta) \sin^4 \theta  d\theta 
\end{equation}

\begin{figure}[!th]
    \centering
    \includegraphics[width=0.9\linewidth]{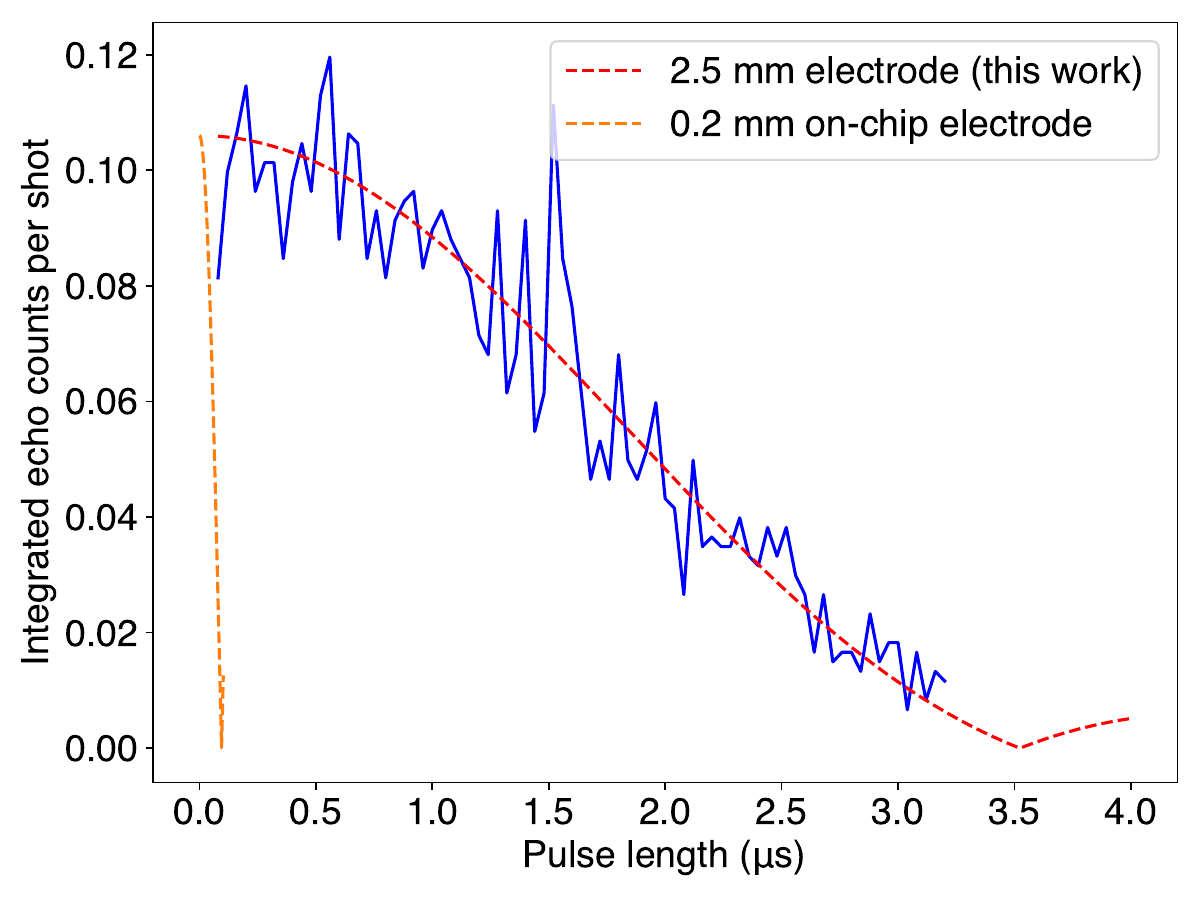}
    \caption{Expected complete extinction of echo using on-chip electrodes.}
    \label{figs6}
\end{figure}

We note that the expression here assumes that the film is polycrystalline with a uniform distribution of the angle of the Er dipole; however, the film may have some degree of structuring which could not be accounted for given the lack of additional information. Fig.~\ref{figs6} below compares the estimated echo suppression using on-chip electrodes separated by 0.2~mm. The $\sin^4 \theta$ term in the expression above changes to $\cos^4 \theta$ for on-chip electrodes with in-plane DC electric field (vs out-of-plane assumed in the derivation above):
\begin{equation}
    A = A_0' \int_0^{\pi/2}  \cos (2 \pi k E t_{pulse} \cos \theta) \cos^4 \theta  d\theta 
\end{equation}
With 40~V applied across on-chip electrodes separated by 0.2~mm, we anticipate electric fields of 500~V/cm, and therefore complete echo suppression in 95~ns (orange lines Fig.~\ref{figs6}) vs 3.5~\textmu s long pulses required here.

\section{Photon echo coherence time at 3.8~K compared to 10~mK}
\begin{figure}[!th]
    \centering
    \includegraphics[width=0.99\linewidth]{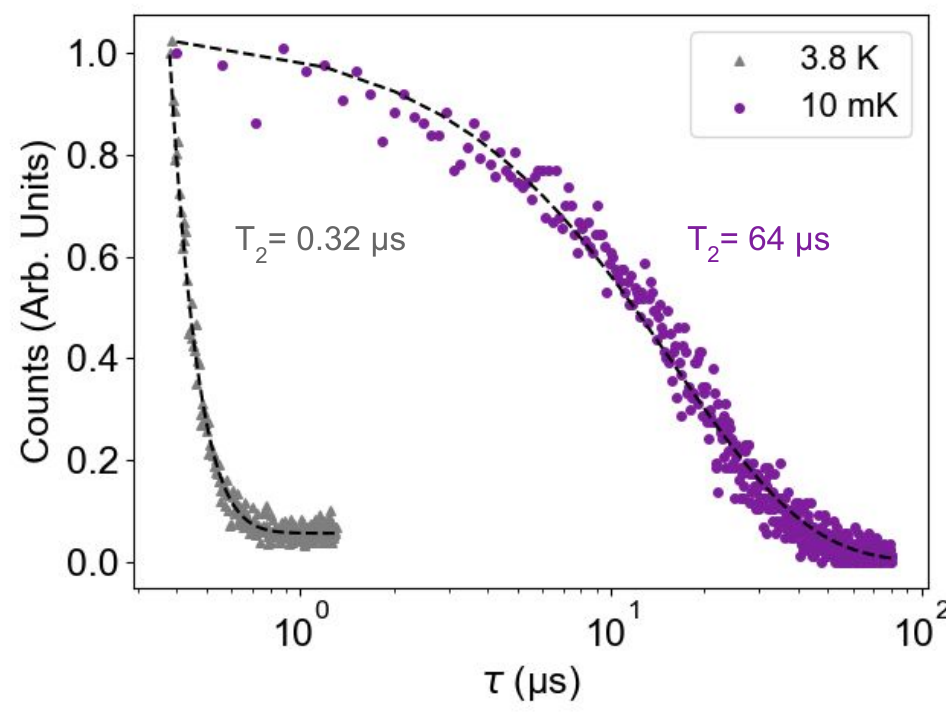}
    \caption{Photon echo homogeneous linewidth for chip3h measured at 3.8~K and 10~mK.}
    \label{figs7}
\end{figure}

Fig.~\ref{figs7} shows the optical coherence time of 326~ns (1~MHz homogeneous linewidth) measured at 3.8~K compared to 64~\textmu s measured at 10~mK on the 3-hour annealed chip, both at 433~mT magnetic fields.

\section{Photon echo amplitude modulation due to superhyperfine interaction}
\begin{figure}[!th]
    \centering
    \includegraphics[width=0.9\linewidth]{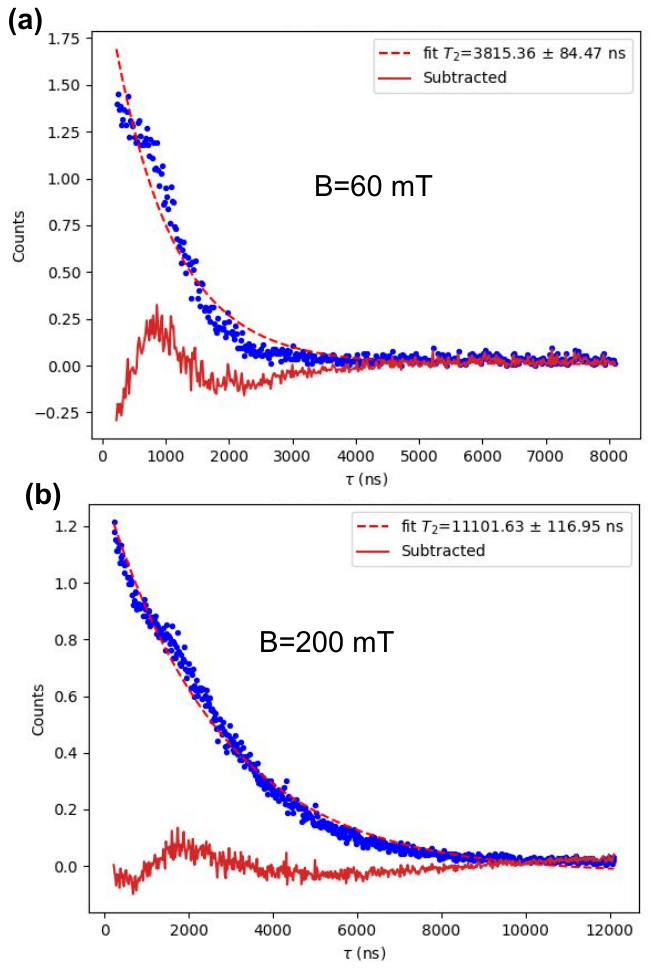}
    \caption{Field-dependent echo envelope modulation due to superhyperfine interaction with nuclear spins for (a) low field and (b) high field transition.}
    \label{figs8}
\end{figure}

The echo envelope shows a field-dependent weak amplitude modulation, with an approximate periodicity of O(100~kHz). The modulation is more pronounced for the lower field transition (Fig.~\ref{figs8}(a)) in comparison with the higher field transition (Fig.~\ref{figs8}(b)), which has been attributed to the host nuclear spin bath in systems like YSO \cite{PhysRevB.102.115119}.  

\section{Estimated Optical Depth}
\begin{figure}[!th]
    \centering
    \includegraphics[width=0.99\linewidth]{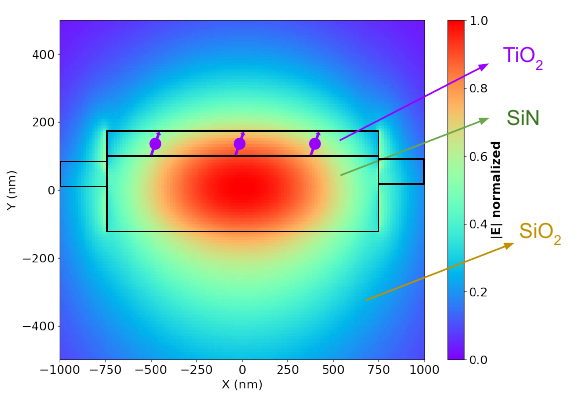}
    \caption{Simulated TE mode profile for SiN waveguide with \TiOtwo .}
    \label{figs9}
\end{figure}

The intrinsic photon echo efficiency on the sample chip1h was estimated from the ratio of the echo area to the input $\pi/2$ pulse area to be $\eta= 1.5 \times 10^{-5}$. This efficiency was obtained after correcting for the sample and setup losses. The echo efficiency is related to the optical depth as \cite{PhysRevA.79.053851}:
\begin{equation}
    \eta = (e^{OD/2} - e^{-OD/2})^2 \approx (OD/4)^2
\end{equation}
\noindent where OD=$\mathrm{\alpha L}$. This allows us to estimate OD $\sim$ 0.015 for the sample. This corresponds to an absorption coefficient $\alpha = 0.03$~cm\textsuperscript{-1}, primarily limited by a low mode confinement of the 50~nm Er-doped layer with the evanescent mode of the silicon nitride waveguide as seen in Fig.~\ref{figs9}. 

An OD of approximately 1 can be reached by a combination of a 3-fold increase in Er density, a 4-fold increase from a thicker (120~nm) \TiOtwo\ layer, and a 10-fold increase from a longer (5~cm) silicon nitride waveguide.

Silicon nitride waveguide losses below 0.5 dB/cm have been reported on the AIM QFlex platform \cite{aim} which will allow using cm-long waveguides to maximize light-matter interaction while incurring minimum waveguide loss. For achieving near-unity quantum memory efficiency, an alternative approach involves the exploration of impedance-matched nanophotonic cavities characterized by a cooperativity of one \cite{PhysRevApplied.12.024062}. We note that ensemble cooperativity values nearing unity have already been reported for nanophotonic cavities in higher-density \TiOtwo\ films \cite{Solomon:24} which makes the cavity-based approach viable.

% \import{SI/}{SI.tex}

\end{document}